\documentclass[12pt]{iopart}
\usepackage[dvips]{epsfig}

\begin{document}
\title[Relativistic entanglement in single-particle]{Relativistic entanglement in single-particle quantum  states using Non-Linear entanglement witnesses }

\author{M.A.Jafarizadeh, M.Mahdian}

\address{ Department of Theoretical Physics and Astrophysics,
University of Tabriz, Tabriz 51664, IRAN}
\ead{jafarizadeh@tabrizu.ac.ir and Mahdian@tabrizu.ac.ir }
\begin{abstract}
In this study, the spin-momentum correlation of one massive spin-$\frac{1}{2}$ and spin-1 particle states, which are made based on projection of a relativistic spin operator into timelike direction is investigated. It is shown that by using Non-Linear entanglement witnesses (NLEWs), the effect of Lorentz transformation would decrease both the amount and the region of entanglement.

\end{abstract}

\maketitle

\section{Introduction:}

Relativistic aspects of quantum entanglement, especially in Einstein, Podolsky and Rosen (EPR) correlations \cite{Einstein} and the Bell inequality \cite{Lee,Caban,Adami,Terashimo,Alsing,Ahn}, have recently attracted much attention. The quantum states in  inertial frame of reference  transform under Wigner rotation \cite{Wigner} via Non-Abelian continuous group SU(2) representation in momentum space. Therefore, under Lorentz transformation, spin becomes entangled with the particle's momentum. Photon polarization qubits behave similarly and linear polarization states of photon will be seen in moving frame in the form of  depolarized states \cite{peres1}.\\ Other aspects of relativistic entanglement are based on spinor formalism, which is designed for four-dimensional space-time \cite{penrose}. In Refs.\cite{ Czachor1,Czachor2,Czachor3,Czachor4,mah22}, Czachor and co-worker have used spinor method and shown that the \emph{helicity}-momentum states, which are the irreducible representation of  Poincar´e group, are the projection of the relativistic spin operator on  timelike  direction in momentum representation.\par
Separability of the quantum states and quantification of entanglement in composite systems are perhaps the most important features of quantum information. The first related criterion for distinguishing entangled states from separable ones, is the positive partial transpose (PPT) criterion, introduced by Peres \cite{A}. Nonetheless, the strongest manner to characterize entanglement is using entanglement witnesses EWs  \cite{HorodeckiE,terhal}. An EW is an observable $\mathcal{W}$ whose expectation value is nonnegative on any separable state, but strictly negative on an entangled state. Recently, there has been an increased interest in the NLEWs because of their improved detection with respect to linear EWs. An NLEW is any bound on nonlinear function of observables which is satisfied by separable states but violated by some entangled states \cite{jaf11,jaf22}. For the first time in Refs.\cite{16,17}, it was actually shown that EW can be expressed as a measure of entanglement. However, the main objective of this paper is to show that NLEWs can also be very helpful  to the quantification of entanglement.\par
In our previous work \cite{mah1}, we studied spin-momentum correlation in single-particle spin-$\frac{1}{2}$ quantum states by using  concurrence and showed that the amount of spin-momentum correlation depends on the  angle between spin and momentum. But here, We quantify the spin-momentum correlation for a  massive  spin half and one relativistic single-particle in $2\otimes2$ (as two-qubit system) and  $2\otimes3$ dimension  Hilbert space using the NLEWs. For simplicity, instead of superposition of momenta, we have used two momentum eigenstates ($p_1$ and $p_2$) and  in 2D momentum subspace, our results suggest the effect of the Lorentz transformation would decrease both the amount and the region of entanglement.\\
This paper is organized as follows:  Sec. II, is devoted
to single-particle  spin half and one quantum states. In Sec. III  the spin-momentum  correlation is calculated in  relativistic single-particle spin half and one mixed states by using the NLEW. Finally, in Sec. IV the results are summarized and conclusions are presented.

\section{Relativistic one massive  spin-$\frac{1}{2}$ and one particle  quantum states in \emph{helicity} basis}

\subsection{ Unitary representation of Poincar´e group}
The Poincar´e group ISO(3,1) is  a semidirect product of SO(3,1) group of Lorentz transformation with the following generators
\begin{equation}P^\mu , J^{\mu\nu}   \hspace{8mm}   \mu,\nu=0,1,2,3\label{eq11}
\end{equation}
where $P^\mu, \mu=1,2,3$ are components of momentum operators, and $P^0=p^0=\sqrt{m^2c^2+\textbf{p}^2}$ is \emph{Hamiltonian}, and $ \textbf{J}=\{J^{23}, J^{31},  J^{12}\}$ are components of angular momentum and  $\textbf{K}=\{J^{10}, J^{20},  J^{30}\}$ are boost operators. Likewise, unitary representation of Poincar´e group can be written in the following form:
\numparts
\begin{eqnarray}
\textbf{K}=-ip^0\nabla_\textbf{p}-\frac{\textbf{p}\times \textbf{s}}{mc+p^0},\\
\textbf{J}=-i\textbf{p}\times\nabla_\textbf{p}+\textbf{s},\\
\textbf{P}=\textbf{p},
\end{eqnarray}
\endnumparts
where $\nabla_\textbf{p}$ is the partial differential as $\frac{\partial}{\partial \textbf{p}}$ and \textbf{s} denotes the finite dimensional angular momentum corresponding to $\textbf{s}^2=s(s+1)$.
We know that the Poincare group is of rank 2, so there are only two independent Casimir invariant operators, which are squared mass and the square of the Pauli-Lubanski vector $W_\mu$  that commute with all generators of the algebra. Let's use
this definition to write down the form of Casimir's operators.
\begin{equation}\label{eq2}P^2=P_\mu P^\mu ,   \hspace{5mm} W^2=W_\mu W^\mu,\end{equation}
the Pauli-Lubanski vector can be written as:
\begin{equation}\label{eq2}W_\mu=\frac{1}{2}\epsilon_{\mu\nu\rho\sigma}J^{\nu\rho}P^\sigma\end{equation}
where $\epsilon_{\mu\nu\rho\sigma}$ is anti symmetry Levi-Civita tensor and takes the value +1 if $\mu\nu\rho\sigma$ is an even permutation, -1 if it is an odd permutation and zero otherwise ( $\epsilon_{0123}=-\epsilon^{0123}=1 $) . It is common to denote the Pauli-Lubanski vector as
\begin{equation}\label{eq3}W^\mu=(W^0,\vec{\textbf{W}})=(\textbf{P}\cdot\textbf{J},P_0\textbf{J}-\textbf{P}\times \textbf{K}).\end{equation}
We can define the popular center-of-mass position operator as
\begin{equation}\label{eq4}\textbf{Q}=\frac{-1}{2}[\frac{1}{P_0}\textbf{K}+\textbf{K}\frac{1}{P_0}].\end{equation}
Then, the orbital angular momentum is
\begin{equation}\label{eq5}\textbf{L}=\textbf{Q}\times \textbf{P}.\end{equation}
Therefore, We can write a relativistic spin operator in the following way
\begin{equation}\label{eq6}\textbf{S}=\textbf{J}-\textbf{L},\end{equation}
and using the Pauli-Lubanski vector, we get
\begin{equation}\label{eq7}\textbf{S}=\frac{\textbf{W}}{P^0}.\end{equation}
So, the spacelike component of Pauli-Lubanski vector is proportional to the relativistic spin operator.
If we choose the timelike vector $b^a=(b^0,0,0,0)$, then after projection of Pauli-Lubanski vector on it, we obtain
\begin{equation}\label{helicity} b^a W_a=b^0 W^0=b^0 \textbf{p}\cdot\textbf{J},\end{equation}
which is helicity. On the other hand, the projection of Eq.(\ref{eq3}) onto the vector $b^a=(0,\textbf{b})$  have eigenvalues as
\begin{equation}\label{spinor1}\lambda_{\pm}= \frac{\pm1}{2}\sqrt{(\textbf{p}\cdot\textbf{b})^2+m^2 b^2}.\end{equation}

If the world-vector $b^a$ could be a null tetrad, i.e., $b^2=b^a b_a=0$,  we get
\begin{equation}\label{spinor2}\lambda_{\pm}=\pm\frac{1}{2}(p\cdot b).\end{equation}

In momentum representation, we can write two energy-momentum world-vectors as a linear combination of the null tetrad and after projection of RSO onto the null tetrad (see Eq. (\ref{spinor2})), we get the following eigenvalue which is similar to eigenvalue  Pauli's matrix, i.e.,  $\sigma_z$ ( for details see \cite{Czachor1} ).
\begin{equation}\label{spinor11}\lambda_{\pm}=\pm\frac{1}{2}.\end{equation}

\subsection{ single-particle spin-$\frac{1}{2}$ quantum state }
Here in this paper, quantum state is made up of a single-particle having two
types of degrees of freedom : momentum \emph{p} and spin $\sigma$.
The former is a continuous variable with Hilbert space of infinite
dimension while the latter is a discrete one with Hilbert space of
finite dimension. For simplicity, instead of
using the superposition of momenta, we use only two momentum
eigenstates ($p_1$ and $p_2$). However, we restrict ourselves to 2D momentum subspace with two eigenstates $p_1$ and $p_2$, so the pure quantum state of such a system can always be written as
\begin{equation}\label{p1}|\psi\rangle=\sum_{i=1}^{2}\sum_{\sigma=0,1} c_{ij}|p_i,\sigma\rangle,\end{equation}
where  $|p_{1(2)}\rangle$ and $|\sigma\rangle$   are  eigenstates of momentum and spin operators, respectively. $c_{ij}$ are complex coefficients such that $\sum
_{i,j}|c_{ij}|^2=1 $.\\
A bipartite quantum mixed state is defined as a convex combination of bipartite pure states (\ref{p1}), i.e.,
\begin{equation}\label{ro1}\rho_{\frac{1}{2}}=\sum_{i=1}^4 P_i
|\psi_{i}\rangle \langle\psi_{i}|,\end{equation}
where $P_i\geq 0$, $\sum _{i} P_i=1 .$ The subscript `` $\frac{1}{2}$ '' refers to the spin-$\frac{1}{2}$ and $|\psi_i\rangle$ $(i=1,2,3,4)$ as four orthogonal maximal entangled Bell states (\emph{BD})  belong to
 the product space $\mathcal{H}_p \otimes \mathcal{H}_\sigma$ and are well-known as
\numparts
\begin{eqnarray}
|\psi_1\rangle=\frac{1}{\sqrt{2}}(|p_{1},0\rangle+|p_{2},1\rangle) ,\\
|\psi_2\rangle=\frac{1}{\sqrt{2}}(|p_{1},0\rangle-|p_{2},1\rangle) ,\\
|\psi_3\rangle=\frac{1}{\sqrt{2}}(|p_{2},0\rangle+|p_{1},1\rangle) ,\\
|\psi_4\rangle=\frac{1}{\sqrt{2}}(|p_{2},0\rangle-|p_{1},1\rangle).\end{eqnarray}
\endnumparts
in terms of momentum and spin states. The kets $|0\rangle$ and $|1\rangle$ are the eigenvectors of spin operator $\sigma_z$.
We assumed that spin and momentum are parallel in the z-direction and in this case, the single-particle spin-$\frac{1}{2}$ state can be considered as a two-qubit system. For an observer in another
reference frame $S^\prime$, is described by an arbitrary boost $\Lambda$
in the x-direction. The transformed \emph{BD} states are given by (see Appendix A),
\begin{equation}\label{wig}|\psi_{i}\rangle\longrightarrow U(\Lambda)|\psi_{i}\rangle=\sqrt{\frac{(\Lambda
p)^0}{p^0}}\sum_{\sigma^\prime} D_{\sigma^\prime
\sigma}(W(\Lambda,p))|\psi_{i}\rangle ,\end{equation}
where $U(\Lambda)$ is a unitary representation of Lorentz transformation. It can be calculated that $|\psi_{i}\rangle$ will be orthogonal after Lorentz transformation, i. e.,
\begin{equation}\label{ort}\langle\Lambda \psi_{i}|\Lambda \psi_{j}\rangle=\delta_{ij}.\end{equation}
 The \emph{BD} density matrix (\ref{ro1}), which describes the state of the single-particle at non-relativistic frame, is exchanged to the density matrix $\rho^{\Lambda}_{\frac{1}{2}}$ after Lorentz transformation, i.e.,
\begin{equation}\label{rro1}\rho_{\frac{1}{2}}\longrightarrow U(\Lambda)\rho_{\frac{1}{2}}U(\Lambda)^\dag ,\end{equation}
therefore
\begin{equation}\label{rro}
\rho^{\Lambda}_{\frac{1}{2}}=U(\Lambda)\rho_{\frac{1}{2}}U(\Lambda)^\dag=\sum_{i=1}^{4}
P_{i}|\Lambda \psi_{i}\rangle\langle\Lambda \psi_{i}| .
\end{equation}

\subsection{single-particle spin-1 quantum state}

We assumed that for spin one, the particle moves with two momentum eigen-state $p_1$ and $p_2$ along the y-axis. For an observer in another reference frame $S^\prime$ described by an arbitrary boost $\Lambda$ given by the velocity $\vec{v^\prime} $ in the z-direction, we have
\begin{equation}\label{spi1}\textbf{p}_{1(2)}=(0,\pm p_{1(2)},0,\emph{E})=(0, m\sqrt{{\gamma^\prime_{1(2)}}^2-1},0,m\gamma^\prime_{1(2)}),\end{equation}
where
\begin{equation}\label{spi2}\emph{E}^2-p^2_{1(2)}=m^2 ,\gamma^\prime_{1(2)}=\sqrt{\frac{p^2_{1(2)}+m^2}{m^2}} \ \ ,\bar{\gamma^\prime}=\frac{1}{\sqrt{1-\beta^2}},\beta=\frac{v^\prime}{c} .\end{equation}
According to Ref.\cite{weinberg}, the Wigner representation for spin $j$ is calculated as follows:\\
\begin{equation}\label{wr}
\mathcal{D}_{\sigma^\prime,\sigma}^{j}(W(\Lambda,p))=e^{\frac{i\theta}{2}(\delta_{\sigma^\prime,\sigma+1}\sqrt{(j-\sigma)(j+\sigma+1)}+\delta_{\sigma^\prime,\sigma-1}\sqrt{(j+\sigma)(j-\sigma+1)})},
\end{equation}
where  $\theta$ is angle around the x-axis.
After some mathematical manipulations for spin one we get\\
\begin{equation}\label{wr1}
\begin{array}{ccc}
  \mathcal{D}_{\sigma^\prime,\sigma}^{(\emph{j} =1)}(W(\Lambda,p_{1(2)}))=\left(\begin{array}{ccc}
  \frac{\cos{\theta_{1(2)}}+1}{2} & \frac{i \sin{\theta_{1(2)}}}{\sqrt{2}} & \frac{\cos{\theta_{1(2)}}-1}{2} \\
  \frac{i\sin{\theta_{1(2)}}}{\sqrt{2}} & \cos{\theta_{1(2)}} & \frac{i\sin{\theta_{1(2)}}}{\sqrt{2}} \\
  \frac{\cos{\theta_{1(2)}}-1}{2} & \frac{i\sin{\theta_{1(2)}}}{\sqrt{2}} & \frac{\cos{\theta_{1(2)}}+1}{2} \\
\end{array}
\right)\\
\end{array},
\end{equation}
for simplicity, assume that $p_1=-p_2=p$, then  $\theta_1=-\theta_2=\theta$ .\\
In order to consider single spin one particle mixed state under Lorentz transformation
we define the following density matrix in the rest frame as follows:
\begin{equation}\label{ro5}\rho_1=x
|\psi_5\rangle\langle\psi_5|+y
|\psi_6\rangle\langle\psi_6|+\frac{1-x-y}{6}I, x+y\leq1 ,\end{equation}
The subscript ``1''refers to the spin one and $|\psi_{5(6)}\rangle$ are maximally entangled pure states, given by
\numparts
\begin{eqnarray}
|\psi_5\rangle=\frac{1}{\sqrt{2}}(|p_1,-1\rangle+|p_2,1\rangle),\\
|\psi_6\rangle=\frac{1}{\sqrt{2}}(|p_1,-1\rangle-|p_2,1\rangle),
\end{eqnarray}
\endnumparts
where $|p_{1(2)}\rangle$  are two momentum eigen states of particle and $\{|\pm1\rangle,|0\rangle\}$  are the three component spin ones as
\begin{equation}\label{pures1}|1\rangle=\left(\begin{array}{c}1 \\
0\\0\end{array}\right) , |0\rangle=\left(\begin{array}{c}0 \\
1\\0\end{array}\right), |-1\rangle=\left(\begin{array}{c}0 \\
0\\1\end{array}\right). \end{equation}
By using (\ref{wr1}), we can obtain the relativistic density matrix (\ref{ro5}) as

\begin{equation}\label{rosp1}\rho\longrightarrow \rho_1^{\Lambda}=U(\Lambda)\rho_1 U^\dag(\Lambda).\end{equation}

\section{ Measure of entanglement of single-particle states using NLEW }

 \subsection{ Entanglement Witnesses}

An entanglement witness $\mathcal{W}$ is an observable that reveals the entanglement of some entangled
state $\rho$, i.e., $\mathcal{W}$ is such that $Tr(\mathcal{W}\cdot \rho_s)\geq0$ for all separable $\rho_s$,
but $Tr(\mathcal{W}\cdot \rho_{en})<0$. The existence of an EW for any entangled state is a direct consequence of Hahn-Banach theorem \cite{Hahn} and the space of separable density operators is convex and closed. Geometrically,
EWs can be viewed as hyper planes that separate some
entangled states from the set of separable states and, hyper plane
indicated as a line corresponds to the state with $Tr[\mathcal{W}\cdot\rho_s]=0$.
According to Refs.\cite{Lewenstein,Lewenstein2}, an EW will be optimal if, for
all positive operators P and $\epsilon>0$, the operator
\begin{equation}\label{witne1}\mathcal{W}^\prime=(1+\epsilon)\mathcal{W}-\epsilon P,\end{equation}
is not an EW.\\
When talking about EW's, one has to take an important point into consideration: the so-called decomposable
EWs (DEW), which can be written as
\begin{equation}\label{decom}
\mathcal{W}=\mathcal{P}+Q_{1}^{T_{A}},
 \quad\quad \mathcal{P},Q_{1}\geq0,
\end{equation}
where the operator $Q_{1}$ is positive semidefinite. It can
be easily verified that such witnesses cannot detect any
bound entangled states. $\mathcal{W}$ is non-decomposable EW if it can not be
put in the form (\ref{decom}) (for more details see
\cite{Doherty3}).
One should notice that only non-decomposable EWs can
detect PPT entangled states.

\subsection{ Measure of entanglement of single spin-$\frac{1}{2}$ particle using NLEW }

According to Ref. \cite{jafa1,mah22} we present an NLEW for a bipartite system ${\cal{H}}_{2}\otimes {\cal{H}}_{2}$ as follows
\begin{equation}\label{bpco1}
   \mathcal{W}_{2\otimes2}=I_{2}\otimes I_{2}+\sum_{i,j=1}^{4}A_{i,j} \hat{\mathcal{O}_{i}}\otimes \hat{\mathcal{O}_{j}},
\end{equation}
where $I_{2}$ is a $2\times 2$ identity matrix, $A_{i,j}$ are some parameters, and $\mathcal{O}_{i}$s  are  Hermitian operators from first ( second ) party Hilbert space as following
\numparts
\begin{eqnarray}
\label{ho2}\mathcal{O}_1=|0\rangle\langle0| ,\\
 \mathcal{O}_2=|1\rangle\langle1|,\\
 \mathcal{O}_3=\frac{1}{\sqrt{2}}(\mathcal{O}_1+\mathcal{O}_2),\\
 \mathcal{O}_4=\frac{-i}{\sqrt{2}}(\mathcal{O}_1-\mathcal{O}_2).\label{ho1}
\end{eqnarray}
\endnumparts
We introduce the maps  for any separable state $\rho_{s}$ which map the convex set of separable states to a bounded convex region  named as feasible region (FR). Then, recalling the definition of an EW, we imposed the first condition which is the problem of the minimization of expectation
values of witness operators $\mathcal{W}$ with respect to separable states, i.e.,
\begin{equation}\label{mi1}min \ \ Tr(\mathcal{W} \cdot\rho_s)\geq0,\end{equation}
In the second step, for a given $\rho_{ent}$, we imposed the second condition for an EW, $Tr(\mathcal{W}\cdot \rho_{ent})< 0$. Now the objective function ( which will be minimized ) is $Tr(\mathcal{W}\cdot \rho_{ent})$, and the inequality constraints come from the first step solution. So, this problem can be written as a convex optimization problem. As mentioned above, we will use two steps towards  finding the parameters $\mathcal{A}_{i,j}$ for the density matrix and  fully characterize NLEWs based on exact convex optimization method for single spin-$\frac{1}{2}$ quantum mixed state.
Now we define $\tilde{\rho_{\frac{1}{2}}}$ as follows:

\begin{equation}\label{RoTilta}
    \tilde{\rho_{\frac{1}{2}}}_{i,j}=Tr(\rho_{\frac{1}{2}}\cdot  \mathcal{O}_{i} \otimes \mathcal{O}_{j}),
\end{equation}
Obviously, these components are associated with components of witness matrix, i.e., $\mathcal{A}_{\mu,\nu}$.
If  matrix $Z$ is a symmetric matrix (i.e. $Z^{t}=Z$) then
$ \tilde{\rho_{\frac{1}{2}}}=-2\mathcal{A}Z$ or
\begin{equation}\label{Z}
    Z=\frac{1}{2}\mathcal{A}^{t} \tilde{\rho_{\frac{1}{2}}}=\frac{1}{2}( \tilde{\rho_{\frac{1}{2}}} ^{t} \tilde{\rho_{\frac{1}{2}}})^{\frac{1}{2}},
\end{equation}
where
\begin{equation}\label{A}
    \mathcal{A}=-\frac{1}{2} \tilde{\rho_{\frac{1}{2}}} Z^{-1}.
\end{equation}
 So, after some calculation we get

\numparts
\begin{eqnarray}\label{a1}
A_{11}=A_{22}=\frac{1}{2}(\frac{-P_1-P_2+P_3+P_4}{\sqrt{(P_1+P_2-P_3-P_4)^2}}-1),\\ A_{12}=A_{21}=\frac{1}{2}(\frac{P_1+P_2-P_3-P_4}{\sqrt{(P_1+P_2-P_3-P_4)^2}}-1), \\
A_{33}=\frac{-P_1+P_2-P_3+P_4}{\sqrt{(P_1-P_2+P_3-P_4)^2}}),\\
A_{44}=\frac{P_1-P_2-P_3+P_4}{\sqrt{(P_1-P_2-P_3+P_4)^2}}, \end{eqnarray}
\endnumparts
and other parameters are zero. Then, by using (\ref{bpco1}) the NLEW can be written as\\
\begin{eqnarray}\label{w1}
\fl \mathcal{W}_{2\otimes2}=I_2\otimes I_2+A_{11}(\mathcal{O}_1\otimes \mathcal{O}_1+\mathcal{O}_2\otimes \mathcal{O}_2)+A_{12}(\mathcal{O}_1\otimes \mathcal{O}_2+\mathcal{O}_2\otimes \mathcal{O}_1)\nonumber\\+A_{33}\mathcal{O}_3\otimes \mathcal{O}_3+A_{44}\mathcal{O}_4\otimes \mathcal{O}_4.\end{eqnarray}\\
Then measuring the observable $\mathcal{W}_{2\otimes2}$ of $\rho_\frac{1}{2}$ gives a good estimate of the expected value of $\mathcal{W}_{2\otimes2}$,\\
$$Tr[\mathcal{W}_{2\otimes2}\cdot\rho_\frac{1}{2}]=\frac{1}{2}\{1-|P_1+P_2-P_3-P_4|-|P_1-P_2+P_3-P_4|-|P_1-P_2-P_3+P_4|\},$$\\
using  $\sum _{i} P_i=1 $   we obtain
\begin{eqnarray}\label{ww11}Tr[\mathcal{W}_{2\otimes2}\cdot\rho_\frac{1}{2}]=(1-2P_i).\end{eqnarray}
We use the concurrence to measure the entanglement between spin and momentum of single particle. It is defined as
$$C(\rho)=max(0,\nu_1-\nu_2-\nu_3-\nu_4),$$

where the quantities  $\nu_1\geq\nu_2\geq\nu_3\geq\nu_4$ in decreasing order are the square roots of the eigenvalues of the matrix
$$ R=\rho (\sigma_y\otimes \sigma_y)\rho^{\ast}(\sigma_y\otimes\sigma_y),$$

for Bell-diagonal  density matrix of (\ref{ro1}) after some calculation we get
\begin{eqnarray}\label{ww22}C(\rho_\frac{1}{2})=(2P_i-1),\end{eqnarray}\\

and Eq.(\ref{ww11}), we obtain
\begin{eqnarray}\label{ww1}C(\rho_\frac{1}{2})=-Tr[\mathcal{W}_{2\otimes2}\cdot\rho_\frac{1}{2}],\end{eqnarray}\\

where $Tr[\mathcal{W}_{2\otimes2}\cdot\rho_\frac{1}{2}]$ is zero when $P_i=\frac{1}{2}$ and exactly coincides
with the concurrence of two-qubit Bell-diagonal mixed state (\ref{ro1}).\\
Moreover, we have made the NLEW for relativistic density matrix (\ref{rro}) and taking the trace we obtain:\\
\begin{equation}\label{MinLag}
    Tr[\mathcal{W}_{2\otimes2}\cdot \rho^{\Lambda}_\frac{1}{2}]=A-\frac{1}{4}(\sqrt{B-4\sqrt{C}}+\sqrt{B+4\sqrt{C}}),
\end{equation}
where \\
\numparts
\begin{eqnarray}\label{a231}
\hspace{-18mm}A=1-\frac{1}{2}\{\sqrt{(P_1+P_2-P_3-P_4)^2}+\sqrt{(P_1-P_2-P_3+P_4)^2}\}\cos{\Omega_p},\\
\hspace{-18mm}B=8(P_1P_3+P_2P_4)\cos^2{\Omega_p}-2(P_1^2+P_2^2+P_3^2+P_4^2)(-3+\cos{2\Omega_p}),\\
\hspace{-18mm}C=4(P_1+P_3)^2(P_2+P_4)^2+4((P_1^2-P_3^2)^2+(P_2^2+P_4^2)^2)\sin^2{\Omega_p}+\nonumber\\4(P_1-P_3)^2(P_2-P_4)^2\sin^42{\Omega_p}.
\end{eqnarray}
\endnumparts
We suppose that the Wigner angle $\Omega_p$ is defined by (see Appendix A)
\begin{equation}\label{ang1}\Omega_{p_1}=\Omega_{p_2}=\Omega_p.\end{equation}
After some calculations and using (\ref{ww11}), we can obtain
\begin{equation}\label{ris}Tr[\mathcal{W}_{2\otimes2}\cdot\rho_\frac{1}{2}]\leq Tr[\mathcal{W}_{2\otimes2}\cdot \rho^{\Lambda}_\frac{1}{2}] \Rightarrow C(\rho_\frac{1}{2}(\Omega_p=0))\geq C(\rho^{\Lambda}_\frac{1}{2}(\Omega_p)).\end{equation}
This result indicates that  both the amount and region of entanglement decrease(see Figure.2).\\

\subsection{Measure of entanglement of single-particle spin-one mixed state using NLEW }
We know that positive partial transpose (PPT) criterion is a necessary and sufficient
condition for determining entangled states living in Hilbert spaces $H_2\otimes H_2$ and $H_2\otimes H_3$. But, with using the NLEW we want to measure of entanglement of bipartite system in $H_2\otimes H_3$ Hilbert space.

In this section, we want to calculate the geometric measure of entanglement of single spin one relativistic particle mixed states in the
 $2\otimes3$ Hilbert space which are spanned by the following basis vectors
\begin{equation}\label{purspin1}\{|p_{1(2)\rangle}\otimes |\pm1\rangle,|p_{1(2)}\rangle\otimes |0\rangle\},\end{equation}
We consider the density matrix of (\ref{ro5}) and obtain the FR by imposing PPT conditions with respect to each party. The PPT condition was applied on the first party and the eigenvalues of  $\rho^T$ in the rest frame (i.e., $\theta=0$) are given by\\
\numparts
\begin{eqnarray}\label{fes1}
\lambda_1=\frac{1}{6}(1+2x-4y),\\
\lambda_2=\frac{1}{6}(1-x-y),\\
\lambda_3=\frac{1}{6}(1-4x+2y),\\
\lambda_4=\frac{1}{6}(1+2x+2y),
\end{eqnarray}
\endnumparts
The positivity of density matrix (\ref{ro5}) imposes the following constraints on the parameters

\numparts
\begin{eqnarray}\label{ppt1}
\lambda_1=\frac{1}{6}(1-x-y),\\
\lambda_2=\frac{1}{6}(1+5x-y),\\
 \lambda_3=\frac{1}{6}(1-x+5y),
 \end{eqnarray}
\endnumparts

we present an EW for mixed state (\ref{ro5}) in ${\cal{H}}_{2}\otimes {\cal{H}}_{3}$ Hilbert space as follows

\begin{equation}\label{bpco1}
   \mathcal{W}_{2\otimes3}=I_{2}\otimes I_{3}+\sum_{i=1}^{4} \sum_{j=1}^{9}\mathcal{A}_{i,j} \hat{\mathcal{O}_{i}}\otimes \hat{\mathcal{Q}_{j}},
\end{equation}.\\
where $I_{2}$ and $I_{3}$  are $2\times 2$ and  $3\times 3$  identity matrix, respectively. $\mathcal{A}_{i,j}$s are some parameters, and $\mathcal{O}_{i}$s which have been defined in the previous section  (see \ref{ho2} until \ref{ho1}) and $\mathcal{Q}_{j}$s are Hermitian operators from second party Hilbert space  as following\\

\numparts
\begin{eqnarray}\label{basis1}
\mathcal{Q}_1=|1\rangle\langle1|,\quad \mathcal{Q}_2=|0\rangle\langle0|,\quad \mathcal{Q}_3=|-1\rangle\langle-1|,\\
\mathcal{Q}_4=\frac{1}{\sqrt{2}}(\mathcal{Q}_1+\mathcal{Q}_2),\quad \mathcal{Q}_5=\frac{1}{\sqrt{2}}(\mathcal{Q}_1+\mathcal{Q}_3),\quad \mathcal{Q}_6=\frac{1}{\sqrt{2}}(\mathcal{Q}_2+\mathcal{Q}_3),\\
\mathcal{Q}_7=\frac{i}{\sqrt{2}}(\mathcal{Q}_1-\mathcal{Q}_2),\quad \mathcal{Q}_8=\frac{i}{\sqrt{2}}(\mathcal{Q}_1-\mathcal{Q}_3),\quad \mathcal{Q}_9=\frac{i}{\sqrt{2}}(\mathcal{Q}_2-\mathcal{Q}_3).\end{eqnarray}
\endnumparts
So, after some mathematical manipulations we get\\
\numparts
\begin{eqnarray}\label{as}
\mathcal{A}_{11}=\mathcal{A}_{23}=\frac{-2\sqrt{3}-(\sqrt{3(x+y)^2}-3\sqrt{2+(x+y)^2})}{6\sqrt{2+(x+y)^2}},\\
\mathcal{A}_{12}=\mathcal{A}_{22}=\frac{-1+x+y}{\sqrt{6+3(x+y)^2}},\\
\mathcal{A}_{13}=\mathcal{A}_{21}=\frac{-2\sqrt{3}-(\sqrt{3(x+y)^2}+3\sqrt{2+(x+y)^2})}{6\sqrt{2+(x+y)^2}},\\
\mathcal{A}_{35}=\mathcal{A}_{48}=-1.
\end{eqnarray}
\endnumparts
Obviously, we can see that the $\mathcal{A}^\dag\mathcal{A}=\mathcal{A}\mathcal{A}^\dag\leq 1$ . So we have an EW candidate as
\begin{eqnarray}\label{www1}
\fl \mathcal{W}_{2\otimes3}=I_2\otimes I_3+\mathcal{A}_{11}(\mathcal{O}_1 \otimes \mathcal{Q}_1+\mathcal{O}_2\otimes  \mathcal{Q}_3)+\mathcal{A}_{12}(\mathcal{O}_1\otimes \mathcal{Q}_2+\mathcal{O}_2\otimes  \mathcal{Q}_2)+\nonumber\\ \mathcal{A}_{13}(\mathcal{O}_1\otimes  \mathcal{Q}_3+\mathcal{O}_2\otimes  \mathcal{Q}_1)-(\mathcal{O}_3\otimes Q_5+\mathcal{O}_4\otimes  \mathcal{Q}_8).
\end{eqnarray}.\\
Its expectation values in separable states are all nonnegative
while its expectation value in the state (\ref{ro5}) reads\\
\begin{equation}Tr(\mathcal{W}_{2\otimes3}\cdot\rho_1)=\frac{1}{6}\{6-6\sqrt{(x-y)^2}-3\sqrt{(x+y)^2}-\sqrt{6+3(x+y)^2}\}.\end{equation}
We want to show that for some values of the parameters $x$ and $y$, the amount of entanglement of mixed state (\ref{ro5}) changes . To this aim, we have constructed a  non-linear EW (\ref{www1}). For convenience, we consider point  $x=0 ,y=1$ which is entangled , i.e., $Tr(\mathcal{W}_{2\otimes3}\cdot\rho)=-1.$
So, in another inertial frame$S^\prime$ that moves with velocity $v$ with respect to rest frame with  $\theta=\frac{\pi}{4}$, we have\\
\numparts
\begin{eqnarray}\label{teta1}
\tau_1=\frac{1}{6}(1-x-y),\\
\tau_{2(3)}=\frac{1}{12}(2-2x+4y\pm3\sqrt{3x^2+y^2}),\\
\tau_{3(4)}=\frac{1}{12}(2+4x-2y\pm3\sqrt{x^2+3y^2}),
\end{eqnarray}
\endnumparts\\
where $\tau_i$'s are eigenvalues of the partial transpose $\rho^{\Lambda}_1=U(\Lambda)\rho_1 U(\Lambda)^\dag.$
So, the non-linear EW, $\mathcal{W}_{2\otimes3}$,  can be constructed in a similar way for relativistic density matrix. Its expectation value with respect to relativistic density matrix $\rho^{\Lambda}_1$ is given by\\
\begin{equation}\hspace{-10mm}Tr(\mathcal{W}_{2\otimes3}\cdot\rho^{\Lambda}_1)=1-\frac{\sqrt{3}}{4}\{\sqrt{(x-y)^2}+\sqrt{(x+y)^2}+\sqrt{B-2\sqrt{A}}+\sqrt{B+2\sqrt{A}}\},\end{equation}
where
\numparts
\begin{eqnarray}\label{cof1}
B=4+11x^2-8xy+11y^2,\\
A=4+4x^2(1+x)(5x-1)-4xy(4x+11x^2-9)+\nonumber\\
y^2(97x^2-16x-4)+ 4(4-11x)y^3+20y^4.
\end{eqnarray}
\endnumparts\\
After some calculations one arrives at
\begin{equation}Tr[\mathcal{W}_{2\otimes3}\cdot\rho_1]\leq Tr[\mathcal{W}_{2\otimes3}\cdot\rho^{\Lambda}_1].\end{equation}
For example, for $x=0,y=1$ we obtain $Tr(\mathcal{W}\cdot\rho^{\Lambda}_1)=\frac{-\sqrt{3}}{2},$
this result shows that the measure of entanglement single spin one particle  decreases when the velocity of the observer increases. In Fig. 1, we have shown that the effect of the Lorentz transformation is to increase the region of separable states.(see Figure.1)\\
Moreover, using the geometric measure of entanglement for quantum systems defined in \cite{hilbersh,16}, we consider entangled state (\ref{ro5}) with $x=0$ and $y=1$ from the nearest separable state in the case $x=\frac{1}{2},y=\frac{1}{2}$, which have been obtained using the convex optimization. We have
\begin{equation}\|\rho-\rho_s\|^2=\frac{\cos^2{\theta}}{\sqrt{2}},\end{equation}
and
\begin{equation}Tr[\mathcal{W}\cdot\rho]=\frac{-1}{2}\cos{\theta}\sqrt{6-2\cos{2\theta}},\end{equation}

which is separable in case of $\theta=\frac{\pi}{2}.$

\section{SUMMARY AND CONCLUSION}

In this paper, we have considered  spin-momentum correlation of single spin half and one relativistic particle quantum states by using the NLEW, which in case of spin half  coincides exactly with the concurrence. Then, we have shown that for single-particle quantum states which have been constructed based on two types of degrees of freedom spin and momentum, both the amount and the region of entanglement between spin and momentum decrease under Lorentz transformation,  with respect to the increasing of observer velocity. Likewise, we have obtained the nearest distance bound of separable states from entangled density matrix and shown that it leads to  zero in the ultrarelativistic limit. Finally, a natural question arises as how the previous results would generalize to the case of other mixed states. So, the calculations in this study are intended as a point of
reference for the development of an understanding of the measure of entanglement in critical quantum systems based on NLEW.

\vspace{1cm}\setcounter{section}{0}
\setcounter{equation}{0}
\renewcommand{\theequation}{A-\roman{equation}}
{\Large APPENDIX A}\\

{\bf Wigner representation for spin-$\frac{1}{2}$}

In Ref. \cite{weinberg}, it is shown that the effect of an arbitrary Lorentz
transformation $\Lambda$ unitarily implemented as $U(\Lambda)$ on
single-particle states is
$$U(\Lambda)(|p\rangle\otimes|\sigma\rangle)=\sqrt{\frac{(\Lambda
p)^0}{p^0}}\sum_{\sigma^\prime} D_{\sigma^\prime
\sigma}(W(\Lambda,p))(|\Lambda p\rangle\otimes|\sigma^\prime\rangle)
,$$ where
$$W(\Lambda,p)=L^{-1}(\Lambda p)\Lambda
L(p),$$
 is the Wigner rotation \cite{Wigner}. We can view this Wigner rotation as follows: we perform the Lorentz transformation $L(p)$ on the rest frame to obtain a moving frame 1, followed by a transformation from frame 1 to frame 2 with $\Lambda$. Then we return to the rest frame by further
performing $L^{-1}(\Lambda p)$. This rotation of the local frame of rest is the kinematic effect that causes the Thomas precession. We will consider two reference frames in this work: one is the rest frame S and the other
is the moving frame $S^\prime$ in which a particle whose
four-momentum \emph{p} in S is seen as boosted with the velocity
$\vec{v}$. By setting the boost and particle moving directions in
the rest frame to be $\hat{v}$ with $\hat{e}$ as the normal vector
in the boost direction  and $\hat{p}$, respectively, and
$\hat{n}=\hat{e}\times \hat{p}$, the Wigner representation for
spin-$\frac{1}{2}$ particle is found as \cite{Ahn},
$$D^{\frac{1}{2}}(W(\Lambda,p)=\cos{\frac{\Omega_{\vec{p}}}{2}}+i\sin{\frac{\Omega_{\vec{p}}}{2}}(\vec{\sigma}.\hat{n}),$$
where
$$\cos{\frac{\Omega_{\vec{p}}}{2}}=\frac{\cosh{\frac{\alpha}{2}}\cosh{\frac{\delta}{2}}+\sinh{\frac{\alpha}{2}}\sinh{\frac{\delta}{2}}(\hat{e}.\hat{p})}{\sqrt{[\frac{1}{2}+\frac{1}{2}\cosh{\alpha}\cosh{\delta}+\frac{1}{2}\sinh{\alpha}\sinh{\delta}(\hat{e}.\hat{p})]}}.$$
$$\sin{\frac{\Omega_{\vec{p}}}{2}}\hat{n}=\frac{\sinh{\frac{\alpha}{2}}\sinh{\frac{\delta}{2}}(\hat{e}\times\hat{p})}{\sqrt{[\frac{1}{2}+\frac{1}{2}\cosh{\alpha}\cosh{\delta}+\frac{1}{2}\sinh{\alpha}\sinh{\delta}(\hat{e}.\hat{p})]}}.$$
 and
$$\cosh{\alpha}=\gamma=\frac{1}{\sqrt{1-\beta^2}} ,\cosh{\delta}=\frac{\emph{E}}{m} ,\beta=\frac{v}{c}.$$

\newpage

\begin{figure}
\centering
\includegraphics[width=390 pt]{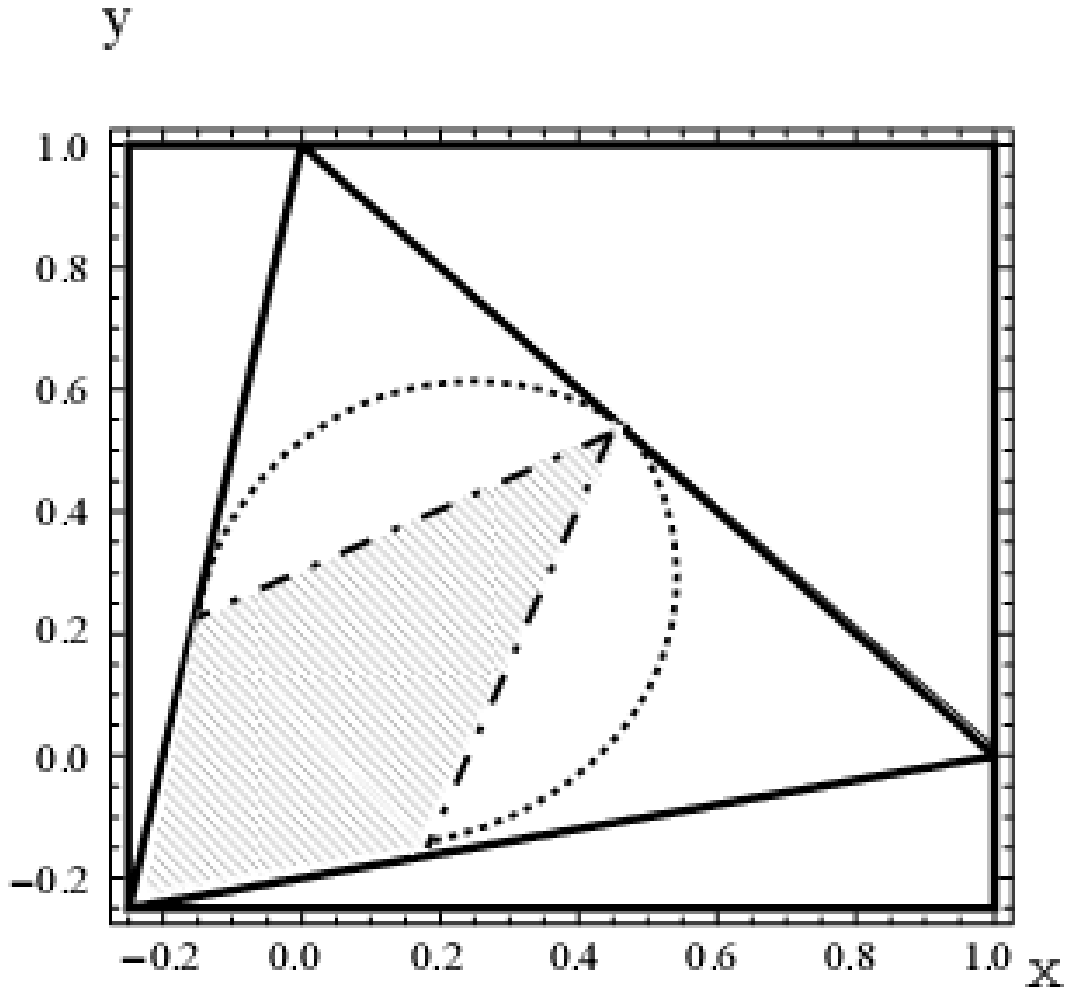}
\caption{}
\label{FIG. 1(a)}
\end{figure}
\textbf{Figure Caption}
\itemize{}
\item Fig. 1: (Color online) Here, the space of the density matrices under Lorentz transformation is shown. The solid line (black triangle)
shows the positivity condition. The hatched area with the dashed lines is the boundary of separability in rest frame(i.e., $\theta=0$). The dotted black curve represents the boundary of separability density matrix under Lorentz transformation when $\theta=\frac{\pi}{4}$. For $\theta=\frac{\pi}{2}$ leads to black triangle and positivity condition in rest frame.

\newpage
\begin{figure}
\centering
\includegraphics[width=390 pt]{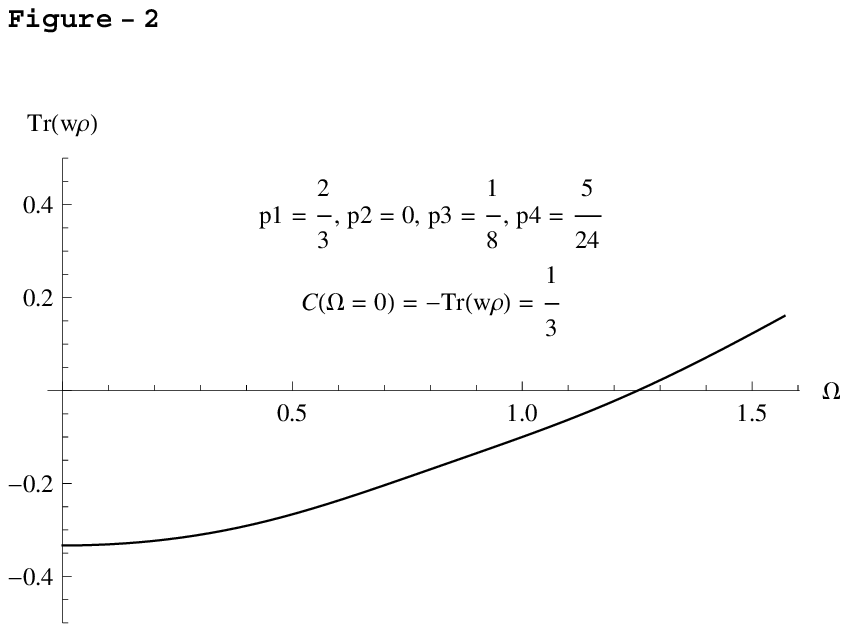}
\caption{}
\label{FIG. 1(b)}
\end{figure}
\textbf{Figure Caption}
\item Fig. 2: The $Tr[\mathcal{W}_{2\otimes2}\cdot \rho^{\Lambda}_\frac{1}{2}]$ versus  Wigner angle $\Omega_p$ . Here $p_1=\frac{2}{3}$, $p_2=0$, $p_3=\frac{1}{8}$, and $p_4=\frac{5}{24}$.

\end{document}